\documentstyle[preprint,aps,eqsecnum,prd]{revtex}
\begin{document}
\draft
\preprint{ITP-SB-94-63}
\title{\large\bf {Top quark inclusive differential distributions}}
\author {N. Kidonakis and J. Smith}
\address{\it Institute for Theoretical Physics, State University of New York at
Stony Brook, \\
Stony Brook, New York 11794-3840}
\date{January 18, 1995}
\maketitle
\begin{abstract}

The inclusive transverse momentum and rapidity distributions
for top quark production at the Fermilab
Tevatron are presented  both in
order $\alpha_s^3$ in QCD and using the
resummation of the leading soft gluon corrections
in all orders of QCD perturbation theory.
The resummed results are uniformly larger than the
$O(\alpha_s^3)$ results for both distributions.
%\vfill
%\newpage
\end{abstract}
\narrowtext
%------------------This is Section 1---------------------------------
\section{Introduction}
%----------------------------------------------------------

At present the top quark has not yet been conclusively
discovered at the Fermilab Tevatron, even though there are events
which look similar to those expected from top quark
decays. The experimental situation is presently
confusing because both the  CDF \cite{CDF} and D0 \cite{D0}
collaborations only have limited statistics. As more events are collected
one expects that the situation will be clarified.

At the Tevatron, the top quark should be mainly
produced through $t\bar{t}$ pair production from the light mass
quarks and gluons in the colliding proton and
antiproton. Both the top quark and the top antiquark
then decay to $(W,b)$ pairs, and
each $W$ boson can decay either hadronically or leptonically.
The $b$-quark becomes an on-mass-shell $B$-hadron which
subsequently decays into leptons and (charmed) hadrons.
A large effort is being made to reconstruct the top quark mass from the
measured particles in the decay, which is complicated by the fact
that the neutrinos are never detected. Also there are
additional jets so it is not clear which ones to choose to
recombine \cite{{OS},{OTS}}. The best channel for this mass
reconstruction is where both $W$ bosons decay
leptonically, one to a $(e,\nu_e)$ pair, the other to a
$(\mu,\nu_{\mu})$ pair (a dilepton event) because the backgrounds
in this channel are small.
When only  a single lepton is detected then it is necessary
to identify the $b$ quark in the decay to remove large backgrounds
from the production of $W +$ jets \cite{Berends}.
In all cases the reconstruction of the particles in the final state involves
both the details of the production of the top quark-antiquark
pair as well as the knowledge of their fragmentation
and decay products. The CDF collaboration \cite{CDF}
have reported two events with dilepton final states, six events
with a single lepton and a $b$-quark identified
by a secondary vertex, and seven single lepton events
with the $b$-quark identified by a semileptonic decay.
The CDF collaboration then constructed a likelihood function for the invariant
mass \cite{Dalitz} and quoted the value
$m_{\rm top} = 174 \pm 10^{+13}_{-12}$ GeV$^/c^2$.
The top quark cross section quoted by the CDF collaboration is
$13.9^{+6.1}_{-4.8}$ pb.
The D0 collaboration \cite{D0} have reported on nine events with
an expected background of
$3.8 \pm 0.9$. If the excess is due to $t\bar t$ production
and if the top quark mass is 180 GeV$/c^2$, then the top quark
cross section is  $8.2 \pm 5.1$ pb.

The values for the top quark production cross section
as a function of the top quark mass used by the CDF
and D0 collaborations contain
both the NLO QCD corrections \cite{{nde1},{betal}}
and an extension to include the  resummation of initial state soft partons to
all orders in perturbation theory \cite {lsn}. A recent
summary of the theoretical predictions has been presented in \cite{ke}.
In the Laenen et al. paper
\cite{lsn} the DIS factorization scheme was used with the
MRSD$\_$ parton distributions \cite{mrs},
the two-loop running coupling constant with five active flavors, and
$\Lambda_{\rm QCD} = 0.152$ GeV.

In the analysis of the decay distributions one needs
knowledge of the inclusive differential distributions
of the heavy quarks in transverse momentum $p_t$ and rapidity $y$.
These distributions are known in NLO \cite{{nde2},{bnmss}}.
What we would like to discuss in this paper is
an updating of the resummation effects on the
inclusive transverse momentum distribution of the top quark.
In the original paper \cite{LSN} it was not known
which mass to choose whereas now
we can assume that the mass is $175$ GeV$/c^2$.
Also we discuss here how
the resummation effects modify the rapidity distribution
of the top quark. Since there have been suggestions of using
the mass and angular distributions in top quark production to
look for physics beyond the standard model \cite{Lane} it is very
important to know the normal QCD predictions for these quantities.

We first summarize what is known on the top quark cross section. If the
top quark mass is 175 GeV$/c^2$ then the dominant production
channel is $q + \bar q \rightarrow t + \bar t$. In lowest order
QCD perturbation theory it contributes
about 90 \% of the total cross section with the reaction
$g + g \rightarrow t + \bar t$ making up the remaining 10\%.
One notes that the NLO corrections
in the $q\bar{q}$ channel are small, whereas those in
the $gg$ channel are more than 80\% .
At this large top quark mass the $qg$ and $\bar{q}g$ channels
give negligible contributions so we do not consider them.
Even though the $gg$ channel contribution is small in Born approximation
it can be significant in NLO due to multiple soft parton radiation.
These large corrections are predominantly from the threshold
region for heavy quark production. It was shown
previously \cite{MSSN} that initial state gluon bremsstrahlung (ISGB)
is responsible for the large corrections at NLO
near threshold.

In \cite{LSN}  the dominant logarithms
from ISGB, which are the cause of the large corrections
near threshold, were carefully examined.
Such logarithms have been studied previously
in Drell-Yan (DY) \cite{DY} production at fixed target energies
(again near threshold) where they are responsible for
correspondingly large corrections.   The
analogy between DY and heavy quark production cross sections was
exploited in \cite{LSN} and
a formula to resum the leading and next-to-leading logarithms
in pQCD to all orders was proposed. Since the contributions due to these
logarithms are positive (when all scales $\mu$ are set equal
to the heavy quark mass $m$),
the effect of summing the higher order corrections
increases the top quark production cross section over that
predicted in $O(\alpha_s^3)$.
This sum, which will be identified as $\sigma_{\rm res}$, depends
on a nonperturbative parameter $\mu_0$. The reason that
a new parameter has to be introduced is that
the resummation is sensitive to the scale at which
pQCD breaks down. As we approach the threshold region
other, nonperturbative, physics plays a role (higher twist,
bound states, etc.) indicated by a dramatic increase in
$\alpha_s$ and in the resummed cross section.
This is commonly called the effect of the infrared renormalon
or Landau pole \cite{Mueller}.
We chose to simply cut off the resummation
at a specific scale $\mu_0$ where
$\Lambda_{\rm QCD} << \mu_0 << m$  since it is not
obvious how to incorporate the nonperturbative effects.
Note that our resummed corrections diverge for small $\mu_0$
but this is {\em not} physical since they should be
joined smoothly onto some nonperturbative prescription
and the total cross section will be finite.
Another way to make it finite would be to avoid the infrared
renormalon by a specific continuation around it, i.e. the
principal value resummation method \cite{{Contop},{Alvero}}.
However, at the moment
our total resummed corrections depend on the parameter
$\mu_0$ for which we can only make a rough estimate.
See \cite{LSN} for more details.

%------------------This is Section 2---------------------------------
\section{Soft Gluon Approximation to the
inclusive distributions}
%----------------------------------------------------------
To make this paper self-contained we list some
relevant formulae.
The partonic processes under discussion will be denoted by
\begin{equation}
i(k_1) + j(k_2) \rightarrow Q(p_1) + \bar Q(p_2),
\end{equation}
where $i,j = g, q, \bar q$. The kinematical variables
\begin{equation}
s = ( k_1+k_2)^2 \quad , \quad t_1 = (k_2-p_2)^2 - m^2 \quad , \quad
u_1 = (k_1- p_2)^2 - m^2\quad ,
\end{equation}
are introduced in the calculation of the corrections
to the single particle inclusive differential distributions
of the heavy (anti)quark. We do not distinguish in the text
between the heavy quark and heavy antiquark since the distributions
are essentially identical in our calculations.
Here $s$ is the square of the
parton-parton c.m. energy and the heavy quark transverse
momentum is given by $p_t= (t_1u_1/s-m^2)^{1/2}$.
The rapidity variable is defined by
$ \exp (2y)  = u_1/t_1$.
The Born approximation differential cross sections can be expressed by
\begin{equation}
s^2\frac{d^2\sigma^{(0)}_{ij}(s,t_1,u_1)}{dt_1 \: du_1} = \delta
(s+t_1+u_1) \sigma^B_{ij}(s,t_1,u_1)\,,
\end{equation}
with
\begin{equation}
\sigma^B_{q\bar q}(s,t_1,u_1) = \pi \alpha_s^2(\mu^2) K_{q\bar{q}}
NC_F \Big[ \frac{t_1^2 + u_1^2}{s^2} + \frac{2m^2}{s}\Big]\,,
\end{equation}
and
\begin{eqnarray}
\sigma^B_{gg}(s,t_1,u_1)& = & 2\pi \alpha_s^2(\mu^2) K_{gg}
NC_F \Big[C_F - C_A \frac{t_1u_1}{s^2}\Big] \nonumber \\ &&
\times\Big[ \frac{t_1}{u_1} + \frac{u_1}{t_1} + \frac{4m^2s}{t_1u_1}
\Big(1 - \frac{m^2s}{t_1u_1}\Big) \Big] \,.
\end{eqnarray}
Here the color factors are
$C_A=N$ and  $C_F=(N^2-1)/(2N)$.
The color average factors are
$K_{q\bar{q}}=N^{-2}$ and $K_{gg}=(N^2-1)^{-2}$.
The parameter $\mu$ denotes the renormalization scale.
In \cite{LSN} the inclusive cross section
was examined near threshold ($s \approx 4m^2$)
where the contributions from the radiation of
soft and collinear gluons are large.
A variable
$s_4 = s+t_1+u_1$ was defined, where $t_1=(k_2-p_2)^2-m^2$
and $u_1=(k_1-p_2)^2-m^2$ are inelastic variables
in the channel $i(k_1)+j(k_2) \rightarrow Q(p_1) + \bar Q(p_2)
+g(k_3)$.
The variable $s_4>0$ now depends on the
four momentum of the extra parton(s) emitted in the reaction.
In the Born approximation there are no additional partons
so $s_4=0$. In \cite{LSN} the NLO contributions were examined
in the soft region (where $s_4 \rightarrow 0$) and it was found
that the dominant contribution to the NLO cross section in this region
had a similar form to the NLO correction in the Drell-Yan
process. As the latter correction is known exactly
in NNLO \cite{Neerven} this correspondence was used to write the differential
cross section in order $\alpha_s^k(\mu^2)$ as follows
\begin{eqnarray}
s^2\frac{d^2\sigma_{ij}^{(k)}(s,t_1,u_1)}{dt_1 \: du_1} &=&
\alpha_s^k(\mu^2) \sum_{l=0}^{2k-1} \Big[\frac{1}{s_4}a_l(\mu^2)
\ln^l\Big(\frac{s_4}{m^2}\Big)\theta(s_4 - \Delta)
  \nonumber \\ &&
+ \frac{1}{l+1} a_l(\mu^2) \ln^{l+1}\Big(\frac{\Delta}{m^2}\Big) \delta(s_4)
\Big] \sigma^B_{ij}(s,t_1,u_1) \,.
\end{eqnarray}
Here a small parameter $\Delta$ has been introduced
to allow us to distinguish between the
soft ($s_4 < \Delta$) and the hard ($s_4> \Delta$)
regions in phase space. The quantities $a_l(\mu^2)$ contain
terms involving the QCD $\beta$-functions and color factors.
The variables $t_1$ and $u_1$ were then mapped onto the variables $s_4$ and
$\cos\theta$, where $\theta$ is the parton-parton
c.m. scattering angle. After explicit integration over the angle
$\theta$, the resulting series was
exponentiated  by the introduction of the $s_4$ variable
into the argument of the running coupling constant.

As noted in the previous paper \cite{LSN} in
addition to the total cross section we can also
derive the resummed heavy (anti)quark inclusive $p_t$ (and below the
$y$) distributions.
The transverse momentum $p_t$ of the heavy quark is related
to our previous variables by
\begin{equation}
t_1 = - \frac{1}{2}\Big\{ s - s_4 -[(s - s_4)^2 - 4s m_t^2]^{1/2}\Big\}\,,
\end{equation}
\begin{equation}
u_1 = - \frac{1}{2}\Big\{ s - s_4 +[(s - s_4)^2 - 4s m_t^2]^{1/2}\Big\}\,,
\end{equation}
with $m_t^2 = m^2 + p_t^2.$ The double differential cross section is
therefore
\begin{equation}
s^2 \frac{d^2\sigma_{ij}(s, t_1, u_1)}{dt_1 \: du_1} =
s[(s - s_4)^2 - 4s m_t^2]^{1/2}
\frac{d^2\sigma_{ij}(s, s_4, p_t^2)}{dp_t^2ds_4} \, ,
\end{equation}
with the boundaries
\begin{equation}
0 < p_t^2 < \frac{s}{4} - m^2\quad , \quad 0 < s_4 < s-2m_t \sqrt{s}\,.
\end{equation}
The $O(\alpha_s^k)$ contribution to the inclusive transverse momentum
distribution $d\sigma_{ij}/dp_t^2$ is given by
\begin{eqnarray}
\frac{d\sigma_{ij}^{(k)}(s,p_t^2)}{dp_t^2} &=&
\frac{2}{s} \alpha_s^k(\mu^2) \sum_{l=0}^{2k-1} a_l(\mu^2)
\int_0^{s-2m_ts^{1/2}}\, ds_4
 \nonumber \\&&
\times\Big\{ \frac{1}{s_4} \ln^l\Big(\frac{s_4}{m^2}\Big) \theta(s_4 - \Delta)
+ \frac{1}{l+1} \ln^{l+1} \Big(\frac{\Delta}{m^2}\Big) \delta(s_4)\Big\}
 \nonumber \\ &&
 \times \frac{1}{[(s-s_4)^2 - 4sm_t^2]^{1/2}} \sigma^B_{ij}(s,s_4,p_t^2)
 \, ,
\end{eqnarray}
where we have inserted an extra factor of two so
that $\int dp_t^2 \: d\sigma/dp_t^2 = \sigma_{\rm tot}$. After some algebra
we can rewrite this result as
\begin{eqnarray}
\frac{d\sigma_{ij}^{(k)}(s,p_t^2)}{dp_t^2} &=&
\alpha_s^k(\mu^2) \sum_{l=0}^{2k-1} a_l(\mu^2)
\Big[\int_0^{s-2m_ts^{1/2}}\, ds_4 \frac{1}{s_4} \ln^l\frac{s_4}{m^2}
 \nonumber \\ &&
\times\Big\{ \frac{d\bar\sigma_{ij}^{(0)}(s,s_4,p_t^2)}{dp_t^2}
 - \frac{d\bar\sigma_{ij}^{(0)}(s,0,p_t^2)}{dp_t^2} \Big\}
  \nonumber \\ &&
+ \frac{1}{l+1} \ln^{l+1}\Big(\frac{s - 2m_t s^{1/2}}{m^2}\Big)
\frac{d\bar\sigma_{ij}^{(0)}(s,0,p_t^2)}{dp_t^2} \Big]\,,
\end{eqnarray}
with the definition
\begin{equation}
\frac{d\bar\sigma_{ij}^{(0)}(s,s_4,p_t^2)}{dp_t^2} =
\frac{2}{s[(s-s_4)^2 - 4s m_t^2]^{1/2}} \sigma_{ij}^B(s,s_4,p_t^2)\,,
\end{equation}
where $d\bar\sigma^{(0)}_{ij}(s,0,p_t^2)/dp_t^2 \equiv
d\sigma^{(0)}_{ij}(s,p_t^2)/dp_t^2 $ again represents the
Born differential $p_t$ distribution. For the $q\bar q$ and $gg$
subprocesses we have the explicit results
\begin{eqnarray}
\frac{d\bar\sigma_{q\bar q}^{(0)}(s,s_4,p_t^2)}{dp_t^2} &=&
2\pi \alpha_s^2(\mu^2) K_{q\bar q} N C_F \frac{1}{s}
\frac{1}{[(s-s_4)^2 -4sm_t^2]^{1/2}}
\nonumber \\ &&
\times \Big[\frac{(s-s_4)^2 - 2sp_t^2}{s^2}\Big] \,,
\end{eqnarray}
and
\begin{eqnarray}
\frac{d\bar\sigma_{gg}^{(0)}(s,s_4,p_t^2)}{dp_t^2} &=&
4\pi \alpha_s^2(\mu^2) K_{gg} N C_F \frac{1}{s}
\frac{1}{[(s-s_4)^2 -4sm_t^2]^{1/2}}
\nonumber \\ &&
\times\Big[C_F - C_A \frac{m_t^2}{s}\Big]
\nonumber \\ &&
\times \Big[\frac{(s-s_4)^2 - 2sm_t^2}{sm_t^2} +
\frac{4m^2}{m_t^2} \Big( 1 -\frac{m^2}{m_t^2}\Big)\Big] \,.
\end{eqnarray}
Since the above formulae are symmetric
under the interchange $t_1 \leftrightarrow u_1$
the heavy quark and heavy antiquark inclusive $p_t$ distributions are
identical. Note that (2.12) is basically the integral of a plus
distribution together with a surface term.

The corresponding formula to (2.12) for the rapidity $y$
of the heavy quark is obtained by using
\begin{equation}
t_1 = - \frac{(s-s_4)}{2}(1 - \tanh y)\,,
\end{equation}
\begin{equation}
u_1 = - \frac{(s-s_4)}{2}(1 + \tanh y)\,.
\end{equation}
The double differential cross section is
therefore
\begin{equation}
s^2 \frac{d^2\sigma_{ij}(s, t_1, u_1)}{dt_1 \: du_1}
=
2 s^2 \frac{\cosh^2y}{s-s_4}
\frac{d^2\sigma_{ij}(s, s_4, y)}{dy \: ds_4}\,,
\end{equation}
with the boundaries
\begin{equation}
- \frac{1}{2}\ln \Big( \frac{1+\beta}{1-\beta}\Big) < y <
  \frac{1}{2}\ln \Big( \frac{1+\beta}{1-\beta}\Big)
\quad , \quad 0 < s_4 < s-2\sqrt{s}m\cosh y\,,
\end{equation}
where $\beta^2 = 1 -4m^2/s$.
The $O(\alpha_s^k)$ contribution to the inclusive rapidity
distribution $d\sigma_{ij}/dy$ is given by
\begin{eqnarray}
\frac{d\sigma_{ij}^{(k)}(s,y)}{dy} &=&
\alpha_s^k(\mu^2) \sum_{l=0}^{2k-1} a_l(\mu^2)
\int_0^{s-2ms^{1/2}\cosh y}\, ds_4
 \nonumber \\&&
\times \Big\{ \frac{1}{s_4} \ln^l\Big(\frac{s_4}{m^2}\Big) \theta(s_4 - \Delta)
+ \frac{1}{l+1} \ln^{l+1}\Big( \frac{\Delta}{m^2}\Big) \delta(s_4)\Big\}
 \nonumber \\ &&
 \times \Big(\frac{s-s_4}{2s^2\cosh^2 y}\Big) \sigma^B_{ij}(s,s_4,y)
 \,.
\end{eqnarray}
After some algebra we can rewrite this result as
\begin{eqnarray}
\frac{d\sigma_{ij}^{(k)}(s,y)}{dy} &=&
\alpha_s^k(\mu^2) \sum_{l=0}^{2k-1} a_l(\mu^2)
\Big[\int_0^{s-2ms^{1/2}\cosh y}\, ds_4 \frac{1}{s_4}
\ln^l\Big(\frac{s_4}{m^2}\Big)
 \nonumber \\ &&
\times\Big\{ \frac{d\bar\sigma_{ij}^{(0)}(s,s_4,y)}{dy}
 - \frac{d\bar\sigma_{ij}^{(0)}(s,0,y)}{dy} \Big\}
  \nonumber \\ &&
+ \frac{1}{l+1} \ln^{l+1}\Big(\frac{s - 2ms^{1/2}\cosh y}{m^2}\Big)
\frac{d\bar\sigma_{ij}^{(0)}(s,0,y)}{dy} \Big]\,,
\end{eqnarray}
with the definition
\begin{equation}
\frac{d\bar\sigma_{ij}^{(0)}(s,s_4,y)}{dy} =
\frac{s-s_4}{2s^2 \cosh^2 y} \, \sigma_{ij}^B(s,s_4,y)\,,
\end{equation}
where $d\bar\sigma^{(0)}_{ij}(s,0,y)/dy \equiv
d\sigma^{(0)}_{ij}(s,y)/dy $ again represents the
Born differential $y$ distribution. For the $q\bar q$ and $gg$
subprocesses we have the explicit formulae
\begin{eqnarray}
\frac{d\bar \sigma_{q\bar q}^{(0)}(s,s_4,y)}{dy} &=&
\pi\alpha_s^2(\mu^2) K_{q\bar q} N C_F
\frac{s-s_4}{2s^2\cosh^2 y}
\nonumber \\ &&
\times \Big[\frac{(s-s_4)^2}{2s^2\cosh^2 y}\Big(
\cosh^2 y + \sinh^2 y\Big) + \frac{2m^2}{s}\Big] \,,
\end{eqnarray}
and
\begin{eqnarray}
\frac{d\bar \sigma_{gg}^{(0)}(s,s_4,y)}{dy} &=&
4\pi \alpha_s^2(\mu^2) K_{gg} N C_F
\frac{s-s_4}{2s^2 \cosh^2 y }
\nonumber \\ &&
\times\Big[C_F - C_A \frac{(s-s_4)^2}{4s^2 \cosh^2 y }\Big]
\times \Big[\cosh^2 y + \sinh^2 y
\nonumber \\ &&
+ \frac{8m^2s \cosh^2 y}{(s-s_4)^2} \Big( 1 -\frac{4m^2s\cosh^2 y}{(s-s_4)^2}
\Big)\Big] \,.
\end{eqnarray}
 Since the above formulae are symmetric under
the interchange $t_1 \leftrightarrow u_1$
the heavy quark and heavy antiquark inclusive $y$-distributions are
identical. Also (2.21) is again of the form of a plus distribution
together with a surface term. Finally, we note that the terms in
(2.12) and (2.21) are all finite.

%------------------This is Section 3---------------------------------
\section{Resummation procedure in parton-parton collisions}
The resummed  contribution to the top quark cross
section can be written as \cite{LSN}
\begin{equation}
s^2\frac{d^2\sigma_{ij}(s,t_1,u_1)}{dt_1 \: du_1}
=\left[\frac{df(s_4/m^2,m^2/\mu^2)}{ds_4}\theta(s_4-\Delta)
+f(\frac{\Delta}{m^2},\frac{m^2}{\mu^2})\delta(s_4) \right]
\sigma_{ij}^B(s,t_1,u_1),
\end{equation}
where
\begin{eqnarray}
f\left(\frac{s_4}{m^2},\frac{m^2}{\mu^2}\right)=
\exp\left\{A\frac{C_{ij}}{\pi}\bar\alpha_s\left(\frac{s_4}{m^2},m^2\right)
\ln^2\frac{s_4}{m^2}\right\}\frac{[s_4/m^2]^{\eta}}{\Gamma(1+\eta)}
\exp(-\eta\gamma_E).
\end{eqnarray}
The straightforward expansion of the exponential plus the change
of the argument in $\bar\alpha_s$ via the renormalization group equations
generates the corresponding leading logarithmic terms written
expicitly in \cite{LSN}.
The scheme dependent $A$ and $\bar\alpha_s$ in the above expression
are given by
\begin{equation}
A=2; \; \; \; \; \; \bar\alpha_s(y,\mu^2)=\alpha_s(y^{2/3}\mu^2)
=\frac{4\pi}{\beta_0 \ln (y^{2/3}\mu^2/\Lambda^2)}\,,
\end{equation}
in the $\overline{\rm MS}$ scheme, and
\begin{equation}
\! \! \! \! \! \! \! \! \! \! \! \! A=1; \; \; \; \; \;
\bar\alpha_s(y,\mu^2)=\alpha_s(y\mu^2)
=\frac{4\pi}{\beta_0 \ln (y\mu^2/\Lambda^2)}\,,
\end{equation}
in the DIS scheme,
where $\beta_0=11/3 \; C_A-2/3 \; n_f$
is the lowest order coefficient of the QCD $\beta$-function.
The color factors $C_{ij}$ are defined by $C_{q\bar{q}}=C_F$ and
$C_{gg}=C_A$, and $\gamma_E$ is the Euler constant.
The quantity $\eta$ is given by
\begin{equation}
\eta=\frac{8C_{ij}}{\beta_0}\ln\left(1+\beta_0
\frac{\alpha_s(\mu^2)}{4\pi}\ln\frac{m^2}{\mu^2}\right)\,.
\end{equation}

Following the procedure in \cite{LSN} for the resummation of the
order $\alpha_s^k$ contributions to the $p_t$ distribution we have
\begin{eqnarray}
\frac{d\sigma_{ij}(s,p_t^2)}{dp_t^2} &=&
\sum_{k=0}^{\infty}
\frac{d\sigma_{ij}^{(k)}(s,p_t^2)}{dp_t^2}
 \nonumber \\ &&
\! \! \! \! \! \! =\int_{s_0}^{s-2m_ts^{1/2}}\, ds_4
\frac{df(s_4/m^2, m^2/\mu^2)}{ds_4}
 \nonumber \\ &&
\times\Big\{ \frac{d\bar\sigma_{ij}^{(0)}(s,s_4,p_t^2)}{dp_t^2}
 - \frac{d\bar\sigma_{ij}^{(0)}(s,0,p_t^2)}{dp_t^2} \Big\}
  \nonumber \\ &&
+ f\Big( \frac{s-2m_ts^{1/2}}{m^2}, \frac{m^2}{\mu^2}\Big)
\frac{d\sigma_{ij}^{(0)}(s,p_t^2)}{dp_t^2} \,.
\end{eqnarray}
Note that we now have cut off the lower limit of the $s_4$ integration
at $s_4=s_0$ because $\bar\alpha_s$ in (3.2) diverges as $s_4 \rightarrow 0$.
This parameter $s_0$ must satisfy
the conditions $0<s_0<s-2m_ts^{1/2}$ and $s_0/m^2 << 1$.
It is convenient to rewrite $s_0$ in terms of the scale $\mu$ as
\begin{mathletters}
\begin{equation}
\frac{s_0}{m^2}=\left(\frac{\mu_0^2}{\mu^2}\right)^{3/2}
(\overline{\rm MS} \: \rm scheme);
\end{equation}
\begin{equation}
\frac{s_0}{m^2}=\frac{\mu_0^2}{\mu^2}
\; \; \; \; \; \; \; \; \; \! (\rm DIS \:scheme).
\end{equation}
\end{mathletters}
Here $\mu_0$ is a nonperturbative parameter  \cite {LSN}
satisfying $\Lambda_{QCD}^2<<\mu_0^2<<\mu^2$.
The derivative of
$f(s_4/m^2,m^2/\mu^2)$ is obtained from
(3.2). It is equal to
\begin{eqnarray}
\frac{df(s_4/m^2,m^2/\mu^2)}{ds_4} &=&
\frac{1}{s_4} \Big\{ 2A \frac{C_{ij}}{\pi} \bar\alpha_s(\frac{s_4}{m^2},
m^2) \ln\frac{s_4}{m^2}+ \eta \Big\}
\nonumber \\ &&
\times \exp\Big\{
 A \frac{C_{ij}}{\pi} \bar\alpha_s(\frac{s_4}{m^2},m^2)\ln^2\frac{s_4}{m^2}
\Big\} \frac{[s_4/m^2]^\eta}{\Gamma(1+\eta)}
\nonumber \\ &&
\times \exp(-\eta\gamma_E)\, ,
\end{eqnarray}
where we have neglected terms which are higher order in
$\bar\alpha_s$.

The analogous formula for the rapidity distribution is
\begin{eqnarray}
\frac{d\sigma_{ij}(s,y)}{dy} &=&
\sum_{k=0}^{\infty}
\frac{d\sigma_{ij}^{(k)}(s,y)}{dy}
 \nonumber \\ &&
\! \!\! \! \! \! =\int_{s_0}^{s-2ms^{1/2}\cosh y}\, ds_4
\frac{df(s_4/m^2, m^2/\mu^2)}{ds_4}
 \nonumber \\ &&
\times\Big\{ \frac{d\bar\sigma_{ij}^{(0)}(s,s_4,y)}{dy}
 - \frac{d\bar\sigma_{ij}^{(0)}(s,0,y)}{dy} \Big\}
  \nonumber \\ &&
+ f\Big( \frac{s-2ms^{1/2}\cosh y}{m^2}, \frac{m^2}{\mu^2}\Big)
\frac{d\sigma_{ij}^{(0)}(s,y)}{dy} \,,
\end{eqnarray}
with the conditions $0<s_0<s-2ms^{1/2}\cosh y$ and $s_0/m^2 << 1$.

%------------------This is Section 4---------------------------------
\section{Numerical Results}

Following the notation in \cite{LSN} the total hadron-hadron
cross section in order $\alpha_s^{k}$ is
%--(4.1)
\begin{equation}
\sigma^{(k)}_H(S,m^2) = \sum_{ij}\int_{4m^2/S}^1
\,d\tau \,\Phi_{ij}(\tau,\mu^2)\, \sigma_{ij}^{(k)}(\tau S,m^2,\mu^2)\,,
\end{equation}
where $S$ is the square of the hadron-hadron c.m. energy and
$i,j$ run over $q,\bar q$ and $g$.
The parton flux $\Phi_{ij}(\tau,\mu^2)$ is defined via
%--(4.2)
\begin{equation}
\Phi_{ij}(\tau,\mu^2) = \int_{\tau}^1\, \frac{dx}{x}
H_{ij}(x,\frac{\tau}{x},\mu^2) \,,
\end{equation}
and $H_{ij}$ is a product of the scale-dependent parton distribution
functions $f^h_i(x,\mu^2)$, where $h$ stands for the hadron which is
the source of the parton $i$
%--(4.3)
\begin{equation}
H_{ij}(x_1, x_2, \mu^2) = f_i^{h_1}(x_1, \mu^2) f_j^{h_2}(x_2,\mu^2)\,.
\end{equation}
The mass factorization scale $\mu$ is chosen to be identical with
the renormalization scale in the running coupling constant.
Since the $p_t$ distribution in hadron-hadron collisions is
not altered by the Lorentz transformation
along the collision axis from the parton-parton c.m. frame,
we can write an analogous formula to (4.1) for the heavy-quark
inclusive differential distribution in $p_t^2$
%--(4.4)
\begin{equation}
\frac{d\sigma^{(k)}_H(S,m^2,p_t^2)}{dp_t^2} = \sum_{ij}\int_{4m_t^2/S}^1
\,d\tau \,\Phi_{ij}(\tau,\mu^2)\, \frac{d\sigma_{ij}^{(k)}
(\tau S,m^2,p_t^2,\mu^2)}{dp_t^2}\,,
\end{equation}
with $m_t^2 = m^2 + p_t^2$. In the case of the all-order
resummed expressions the lower boundaries in (4.1) and (4.4)
have to be modified according to the conditions
$s_0 < s - 2ms^{1/2}$ or $s_0 < s - 2m_ts^{1/2}$ (see above).
Resumming the soft gluon contributions to all orders we obtain
%--(4.5)
\begin{equation}
\sigma^{\rm res }_H(S,m^2) = \sum_{ij}\int_{\tau_0}^1
\,d\tau \,\Phi_{ij}(\tau,\mu^2)\, \sigma_{ij}(\tau S,m^2,\mu^2)\,,
\end{equation}
where $\sigma_{ij}$ is given in (3.24) of \cite{LSN} and
%--(4.6)
\begin{equation}
\tau_0 = \frac{(m+(m^2+s_0)^{1/2})^2}{S}\,,
\end{equation}
with $s_0=m^2(\mu_0^2/\mu^2)^{3/2}$ ($\overline{\rm MS}$ scheme) or
$s_0=m^2(\mu_0^2/\mu^2)$  (DIS scheme) [see (3.7)].
The all-order resummed differential distribution in $p_t^2$ is
%--(4.7)
\begin{equation}
\frac{d\sigma^{\rm res}_H(S,m^2,p_t^2)}{dp_t^2} = \sum_{ij}\int_{\tau_0}^1
\,d\tau \, \Phi_{ij}(\tau,\mu^2) \frac{d\sigma_{ij}(\tau S,m^2,p_t^2,\mu^2)}
{dp_t^2}\,,
\end{equation}
with $d\sigma_{ij}/dp_t^2$ given in (3.6)
and
%--(4.8)
\begin{equation}
\tau_0 = \frac{(m_t+(m_t^2+s_0)^{1/2})^2}{S}\,.
\end{equation}
The corresponding formula to (4.4) for the heavy quark inclusive
differential distribution in $y$ is
\begin{equation}
\frac{d\sigma^{(k)}_H(S,m^2,y)}{dY} = \sum_{ij}\int_{4m^2\cosh^2 y/S}^1
\,d\tau \,\Phi_{ij}(\tau,\mu^2)\, \frac{d\sigma_{ij}^{(k)}
(\tau S,m^2,y,\mu^2)}{dy}\,.
\end{equation}
Order by order in perturbation theory the heavy quark rapidity plots
in the parton-parton c.m. frame show peaks away from $y=0$ (see fig. 7
in \cite{bnmss}). However, upon folding with the partonic densities the
heavy quark rapidities in the hadron-hadron c.m. frame peak near $Y=0$.
Therefore we will assume that the plots for the resummed rapidity distribution
show a similar feature.
The all-order resummed differential distribution in $Y$ is therefore
taken to be
%--(4.7)
\begin{equation}
\frac{d\sigma^{\rm res}_H(S,m^2,Y)}{dY} = \sum_{ij}\int_{\tau_0}^1
\,d\tau \,\Phi_{ij}(\tau,\mu^2)\, \frac{d\sigma_{ij}(\tau S,m^2,y,\mu^2)}
{dy}\,,
\end{equation}
with $d\sigma_{ij}/dy$ given in (3.9)
and
%--(4.8)
\begin{equation}
\tau_0 = \frac{(m\cosh y +(m^2\cosh^2 y+s_0)^{1/2})^2}{S}\,.
\end{equation}
The hadronic heavy quark rapidity $Y$ is related to the
partonic heavy quark rapidity $y$
by
\begin{equation}
Y=y+\frac{1}{2}\ln\frac{x_1}{x_2}\,.
\end{equation}

We now specialize to top quark production at the Fermilab Tevatron
where $\sqrt{S}=1.8$ TeV and choose the top quark mass to be $m=175$
GeV$/c^2$.
In the presentation of our results for the exact, approximate
and resummed hadronic cross sections
we use the MRSD$\_ \:$ parametrization for the parton distributions
\cite{mrs}.
Note that the hadronic results only involve partonic distribution
functions at moderate and large $x$, where there is little difference
between the various sets of parton densities.
We have used the MRSD$\_ \:$ set 34 as given in PDFLIB \cite{PDFLIB} in the
DIS scheme with the number of active light flavors $n_f=5$ and the QCD
scale $\Lambda_5=0.1559$ GeV. We have used the two-loop corrected
running coupling constant as given by PDFLIB.
Since we know the exact $O(\alpha_s^3)$ result, we can make an even
better estimate of the differential distributions by calculating
the perturbation theory improved $p_t$ and $y$ distributions.
We define the improved $p_t$ distribution by
\begin {equation}
\frac{d\sigma_H^{\rm imp}}{dp_t}=\frac{d\sigma_H^{\rm res}}{dp_t}
+\frac{d\sigma_H^{(1)}}{dp_t}\mid _{\rm exact}
-\frac{d\sigma_H^{(1)}}{dp_t}\mid _{\rm app}\,,
\end{equation}
and the improved $Y$ distribution by
\begin {equation}
\frac{d\sigma_H^{\rm imp}}{dY}=\frac{d\sigma_H^{\rm res}}{dY}
+\frac{d\sigma_H^{(1)}}{dY}\mid _{\rm exact}
-\frac{d\sigma_H^{(1)}}{dY}\mid _{\rm app}\,,
\end{equation}
to exploit the fact that $d\sigma_H^{(1)}/dp_t\mid _{\rm exact}$ and
$d\sigma_H^{(1)}/dY\mid _{\rm exact}
$ are known and $d\sigma_H^{(1)}/dp_t\mid _{\rm app}$
and $d\sigma_H^{(1)}/dY\mid _{\rm app}$ are included in
$d\sigma_H^{\rm res}/dp_t$ and $d\sigma_H^{\rm res}/dY$ respectively.
We note that here $d\sigma^{(n)}$ denotes the $O(\alpha_s^{n+2})$
contribution to the differential cross section. Moreover ,
$d\sigma^{(n)}\mid _{\rm exact}$ denotes the exact
calculated differential cross section,
and $d\sigma^{(n)}\mid _{\rm app}$ the approximate one where only the
leading soft gluon corrections are taken into account.

First we present the differential $p_t$ distributions at $\sqrt{S}=1.8$ TeV
for a top quark mass $m = 175$ GeV$/c^2$.
For these plots the mass factorization scale is not everywhere equal to $m$.
We chose $\mu=m$ in $s_0$, $f_k(s_4/m^2\,,m^2/\mu^2)$ and $\bar{\alpha}_s$,
but $\mu=m_t$ in the MRSD$\_$ parton distribution
functions and the running coupling constant $\alpha_s(\mu)$.
 We begin with the results for the $q\bar{q}$ channel in the DIS scheme.
 In fig. 1 we show
 the Born term $d\sigma_H^{(0)}/dp_t$, the first order exact result
$d\sigma_H^{(1)}/dp_t\mid _{\rm exact}$, the first order approximation
$d\sigma_H^{(1)}/dp_t\mid _{\rm app}$, the second
order approximation $d\sigma_H^{(2)}/dp_t\mid _{\rm app}$,
and the resummed result $d\sigma_H^{\rm res}/dp_t$ for $\mu_0=0.05\:m$
and for $\mu_0=0.1\:m$.
These are the same values for $\mu_0$ that were used in \cite{lsn}.
As we decrease $\mu_0$ the differential cross
sections increase. In fig. 2 we show the exact $O(\alpha_s^3)$ result
$d\sigma_H^{(0)}/dp_t+d\sigma_H^{(1)}/dp_t\mid _{\rm exact}$, and,
for comparison, $d\sigma_H^{\rm imp}/dp_t$
for $\mu_0=0.05\:m$ and $\mu_0=0.1\:m$. The resummed differential cross
sections were calculated with the cut $s_4>s_0$ while no such cut was imposed
on the phase space for the individual terms in the perturbation series.
The improved differential distributions are uniformly above the exact
$O(\alpha_s^3)$ result. It is also evident from fig. 2 that the
resummation of the soft gluon contributions to the $p_t$ distribution
modifies the exact $O(\alpha_s^3)$ result only slightly for the values
of $\mu_0$ that have been chosen.

We continue with the results for the $gg$ channel in
the $\overline{\rm MS}$ scheme.
The corresponding plots are given in figures 3 and 4.
In this case the values of $\mu_0$ have been chosen to be
$\mu_0=0.2\:m$ and $\mu_0=0.25\:m$ to correspond to those in \cite{lsn}.
Note that $\mu_0$ need not be the same in the $q\bar{q}$ and $gg$ reactions
because the convergence properties of the QCD perturbation series could be
different in these channels and moreover depend on the factorization scheme.
The first and second order corrections in the $gg$ channel in the
$\overline{\rm MS}$ scheme are larger than the respective ones in the
$q\bar{q}$ channel in the DIS scheme. In fact, for the range of $p_t$
values shown the second-order approximate correction is larger than the
first-order approximation. Hence, the relative difference in magnitude
 between the improved $d\sigma_H^{\rm imp}/dp_t$ and
the exact $O(\alpha_s^3)$ results
is significantly larger than that in the $q\bar{q}$ channel
in the DIS scheme.

We finish our discussion of the differential $p_t$ distributions with the
results of adding the $q\bar{q}$ and $gg$ channels. The plots appear
in figures 5 and 6. It is evident that resummation produces an
enhancement of the exact $O(\alpha_s^3)$ result, with very little change
in shape.

Now we turn to a discussion of the differential $Y$
distributions at $\sqrt{S}=1.8$ TeV
for a top quark mass $m = 175$ GeV$/c^2$.
In this case we set the factorization mass scale equal to $m$ everywhere.
We begin with the results for the $q\bar{q}$ channel in the DIS scheme.
In fig. 7 we show
the Born term $d\sigma_H^{(0)}/dY$, the first order exact result
$d\sigma_H^{(1)}/dY\mid _{\rm exact}$, the first order approximation
$d\sigma_H^{(1)}/dY\mid _{\rm app}$, the second
order approximation $d\sigma_H^{(2)}/dY\mid _{\rm app}$,
and the resummed result $d\sigma_H^{\rm res}/dY$ for $\mu_0=0.05\:m$
and $\mu_0=0.1\:m$. In fig. 8 we show the exact $O(\alpha_s^3)$ result
$d\sigma_H^{(0)}/dY+d\sigma_H^{(1)}/dY\mid _{\rm exact}$, and,
for comparison, $d\sigma_H^{\rm imp}/dY$
for $\mu_0=0.05\:m$ and $\mu_0=0.1\:m$. Again, the resummed differential cross
sections were calculated with the cut $s_4>s_0$ while no such cut was imposed
on the phase space for the individual terms in the perturbation series.
It is also evident from fig. 8 that the
resummation of the soft gluon contributions to the $Y$ distribution
modifies the exact $O(\alpha_s^3)$ result only slightly for the
values of $\mu_0$
that have been chosen.

We continue with the results for the $gg$ channel
in the $\overline{\rm MS}$ scheme.
The corresponding plots are given in figures 9 and 10. Here, the values
of $\mu_0$ are $\mu_0=0.2\:m$ and $\mu_0=0.25\:m$. As in the case of
the $p_t$ distributions, the first and second order
corrections in this channel are larger than the respective ones in the
$q\bar{q}$ channel in the DIS scheme.  For the range of $Y$
values shown the second-order approximate correction is larger than the
first-order approximation. Again, as in the $p_t$ distributions,
the relative difference in magnitude between the improved
$d\sigma_H^{\rm imp}/dY$ and the exact $O(\alpha_s^3)$ results
is significantly larger than that in the $q\bar{q}$ channel
in the DIS scheme.

Finally, we conclude our discussion of the differential
$Y$ distributions by showing the
results of adding the $q\bar{q}$ and $gg$ channels. The plots appear
in figures 11 and 12. Again, it is evident that resummation produces
a non-negligible modification of the exact $O(\alpha_s^3)$ result.
However, the shape of the distribution is unchanged.

We have shown that the resummation of soft gluon radiation
produces a small difference between the perturbation improved distributions
in $p_t$ and $Y$ and the exact $O(\alpha_s^3)$ distributions
in $p_t$ and $Y$ for
the $q\bar{q}$ reaction in the DIS scheme for the values of $\mu_0$ chosen.
However, for the $gg$ channel in the $\overline{\rm MS}$ scheme the
resummation produces a large difference. The difference between the
 perturbation improved and the exact $O(\alpha_s^3)$ distributions depends
on the mass factorization
scheme (DIS or $\overline{\rm MS}$), the factorization scale $\mu$,
as well as the
specific reaction under consideration ($q\bar{q}$ or $gg$).
For a mass $m=175$ GeV$/c^2$ the $gg$ channel is not
as important numerically as
the $q\bar{q}$ channel. However, since the corrections for the $gg$ channel
are quite large, resummation produces a non-negligible difference between
the perturbation improved and the exact
$O(\alpha_s^3)$ distributions when adding the
two channels. However, the shapes of the distributions are essentially
unchanged.

{\bf ACKNOWLEDGEMENTS}

We thank E. Laenen and W. L. van Neerven for helpful discussions.

The work in this paper was supported in part under the
contract NSF 93-09888.
%----------------------------------------------------------

%----------------------------References-------------------------------------
%

\begin{figure}
\caption{The top quark $p_t$ distributions $d\sigma_H^{(k)}/dp_t$ for
the $q\bar{q}$ channel in the DIS scheme for a top quark mass
$m=175$ GeV$/c^2$. Plotted are $d\sigma_H^{(0)}/dp_t$ (upper solid line),
$d\sigma_H^{(1)}/dp_t\mid _{\rm exact}$ (lower solid line),
$d\sigma_H^{(1)}/dp_t\mid _{\rm app}$ (upper dotted line),
$d\sigma_H^{(2)}/dp_t\mid _{\rm app}$ (lower dotted line),
and $d\sigma_H^{\rm res}/dp_t$ ($\mu_0=0.05\:m$ upper dashed line
and $\mu_0=0.1\:m$ lower dashed line).}
\label{fig1}
\end{figure}

\begin{figure}
\caption{The top quark $p_t$ distributions  $d\sigma_H/dp_t$ for the
$q\bar{q}$ channel in the DIS scheme for a top quark mass $m=175$ GeV$/c^2$.
Plotted are $d\sigma_H^{(0)}/dp_t+d\sigma_H^{(1)}/dp_t\mid _{\rm exact}$
(solid line) and $d\sigma_H^{\rm imp}/dp_t$ ($\mu_0=0.05\:m$
upper dashed line and $\mu_0=0.1\:m$ lower dashed line).}
\label{fig2}
\end{figure}

\begin{figure}
\caption{The top quark $p_t$ distributions $d\sigma_H^{(k)}/dp_t$ for
the $gg$ channel in the $\overline{\rm MS}$ scheme for a top quark mass
$m=175$ GeV$/c^2$. Plotted are $d\sigma_H^{(0)}/dp_t$ (upper solid line
at large $p_t$),
$d\sigma_H^{(1)}/dp_t\mid _{\rm exact}$ (lower solid line at large $p_t$),
$d\sigma_H^{(1)}/dp_t\mid _{\rm app}$ (lower dotted line),
$d\sigma_H^{(2)}/dp_t\mid _{\rm app}$ (upper dotted line),
and $d\sigma_H^{\rm res}/dp_t$ ($\mu_0=0.2\:m$ upper dashed line
and $\mu_0=0.25\:m$ lower dashed line).}
\label{fig3}
\end{figure}

\begin{figure}
\caption{The top quark $p_t$ distributions  $d\sigma_H/dp_t$ for the
$gg$ channel in the $\overline{\rm MS}$ scheme for a top
quark mass $m=175$ GeV$/c^2$.
Plotted are $d\sigma_H^{(0)}/dp_t+d\sigma_H^{(1)}/dp_t\mid _{\rm exact}$
(solid line) and $d\sigma_H^{\rm imp}/dp_t$ ($\mu_0=0.2\:m$
upper dashed line and $\mu_0=0.25\:m$ lower dashed line).}
\label{fig4}
\end{figure}

\begin{figure}
\caption{The top quark $p_t$ distributions $d\sigma_H^{(k)}/dp_t$ for
the sum of the $q\bar{q}$ and $gg$ channels for a top quark mass
$m=175$ GeV$/c^2$. Plotted are $d\sigma_H^{(0)}/dp_t$ (upper solid line),
$d\sigma_H^{(1)}/dp_t\mid _{\rm exact}$ (lower solid line),
$d\sigma_H^{(1)}/dp_t\mid _{\rm app}$ (upper dotted line),
$d\sigma_H^{(2)}/dp_t\mid _{\rm app}$ (lower dotted line),
and $d\sigma_H^{\rm res}/dp_t$ (upper
and lower dashed lines).}
\label{fig5}
\end{figure}

\begin{figure}
\caption{The top quark $p_t$ distributions  $d\sigma_H/dp_t$ for the sum
of the $q\bar{q}$ and $gg$ channels for a top quark mass $m=175$ GeV$/c^2$.
Plotted are $d\sigma_H^{(0)}/dp_t+d\sigma_H^{(1)}/dp_t\mid _{\rm exact}$
(solid line) and $d\sigma_H^{\rm imp}/dp_t$
(upper and lower dashed lines).}
\label{fig6}
\end{figure}

\begin{figure}
\caption{The top quark $Y$ distributions $d\sigma_H^{(k)}/dY$ for
the $q\bar{q}$ channel in the DIS scheme for a top quark mass
$m=175$ GeV$/c^2$. Plotted are $d\sigma_H^{(0)}/dY$ (upper solid line),
$d\sigma_H^{(1)}/dY\mid _{\rm exact}$ (lower solid line),
$d\sigma_H^{(1)}/dY\mid _{\rm app}$ (upper dotted line),
$d\sigma_H^{(2)}/dY\mid _{\rm app}$ (lower dotted line),
and $d\sigma_H^{\rm res}/dY$ ($\mu_0=0.05\:m$ upper dashed line
and $\mu_0=0.1\:m$ lower dashed line).}
\label{fig7}
\end{figure}

\begin{figure}
\caption{The top quark $Y$ distributions  $d\sigma_H/dY$ for the
$q\bar{q}$ channel in the DIS scheme for a top quark mass $m=175$ GeV$/c^2$.
Plotted are $d\sigma_H^{(0)}/dY+d\sigma_H^{(1)}/dY\mid _{\rm exact}$
(solid line) and $d\sigma_H^{\rm imp}/dY$ ($\mu_0=0.05\:m$
upper dashed line and $\mu_0=0.1\:m$ lower dashed line).}
\label{fig8}
\end{figure}

\begin{figure}
\caption{The top quark $Y$ distributions $d\sigma_H^{(k)}/dY$ for
the $gg$ channel in the $\overline{\rm MS}$ scheme for a top quark mass
$m=175$ GeV$/c^2$. Plotted are $d\sigma_H^{(0)}/dY$ (upper solid line
at $Y=0$),
$d\sigma_H^{(1)}/dY\mid _{\rm exact}$ (lower solid line at $Y=0$),
$d\sigma_H^{(1)}/dY\mid _{\rm app}$ (lower dotted line),
$d\sigma_H^{(2)}/dY\mid _{\rm app}$ (upper dotted line),
and $d\sigma_H^{\rm res}/dY$ ($\mu_0=0.2\:m$ upper dashed line
and $\mu_0=0.25\:m$ lower dashed line).}
\label{fig9}
\end{figure}

\begin{figure}
\caption{The top quark $Y$ distributions  $d\sigma_H/dY$ for the
$gg$ channel in the $\overline{\rm MS}$ scheme
for a top quark mass $m=175$ GeV$/c^2$.
Plotted are $d\sigma_H^{(0)}/dY+d\sigma_H^{(1)}/dY\mid _{\rm exact}$
(solid line) and $d\sigma_H^{\rm imp}/dY$ ($\mu_0=0.2\:m$
upper dashed line and $\mu_0=0.25\:m$ lower dashed line).}
\label{fig10}
\end{figure}

\begin{figure}
\caption{The top quark $Y$ distributions $d\sigma_H^{(k)}/dY$ for
the sum of the $q\bar{q}$ and $gg$ channels  for a top quark mass
$m=175$ GeV$/c^2$. Plotted are $d\sigma_H^{(0)}/dY$ (upper solid line),
$d\sigma_H^{(1)}/dY\mid _{\rm exact}$ (lower solid line),
$d\sigma_H^{(1)}/dY\mid _{\rm app}$ (upper dotted line),
$d\sigma_H^{(2)}/dY\mid _{\rm app}$ (lower dotted line),
and $d\sigma_H^{\rm res}/dY$ (upper
and lower dashed lines).}
\label{fig11}
\end{figure}

\begin{figure}
\caption{The top quark $Y$ distributions  $d\sigma_H/dY$ for the
sum of the $q\bar{q}$ and $gg$ channels for a top quark mass $m=175$ GeV$/c^2$.
Plotted are $d\sigma_H^{(0)}/dY+d\sigma_H^{(1)}/dY\mid _{\rm exact}$
(solid line) and $d\sigma_H^{\rm imp}/dY$
(upper and lower dashed lines).}
\label{fig12}
\end{figure}

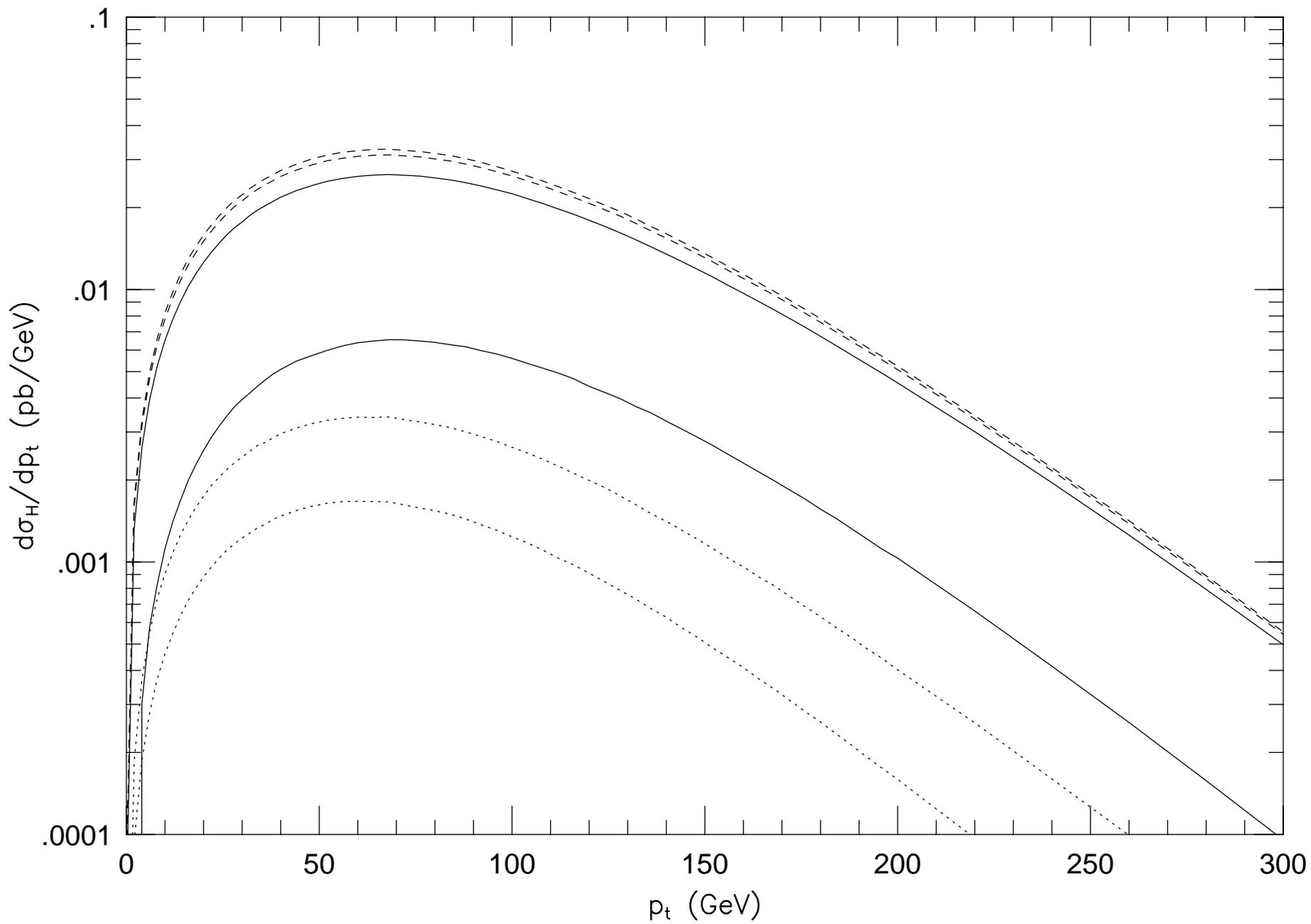
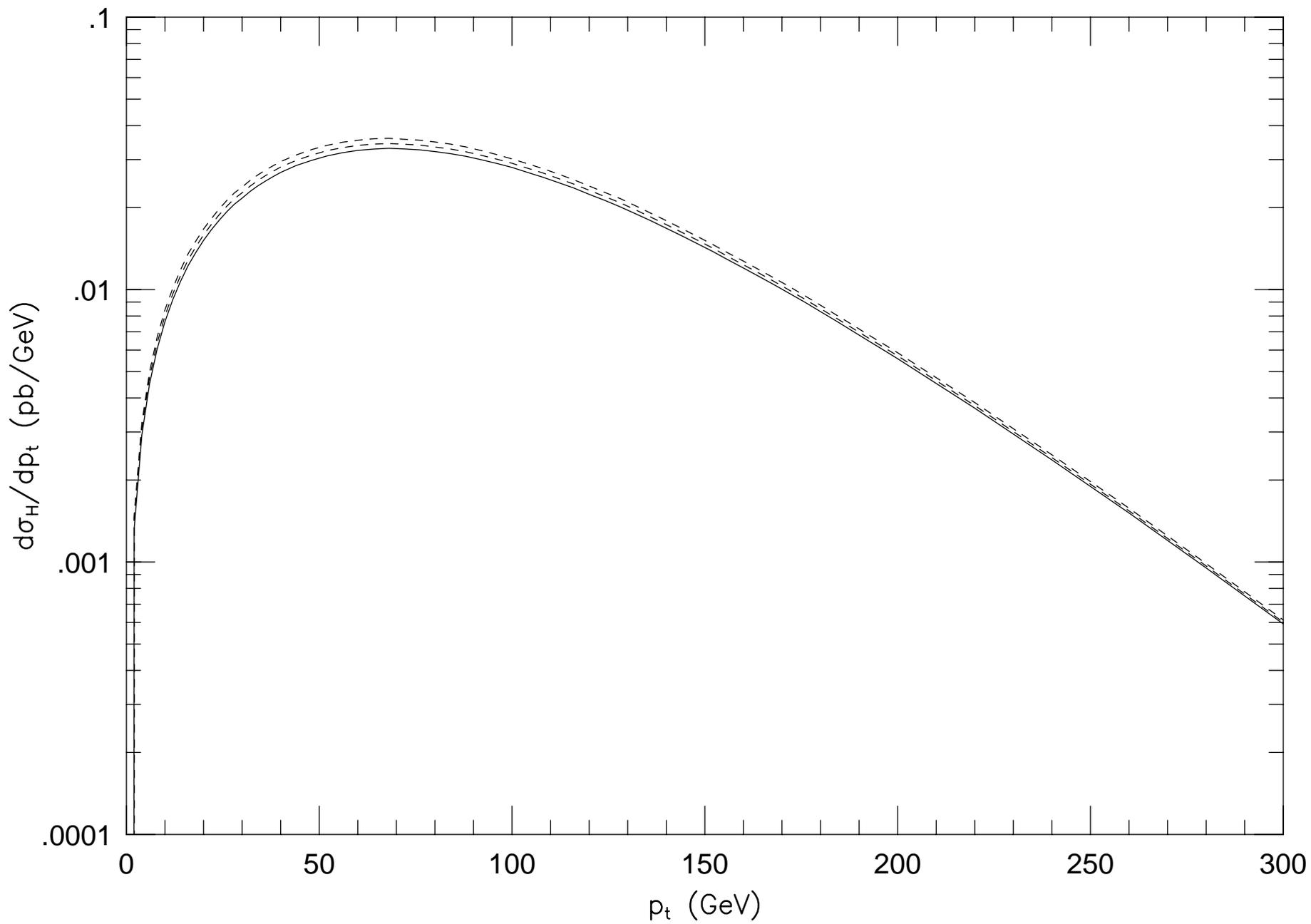
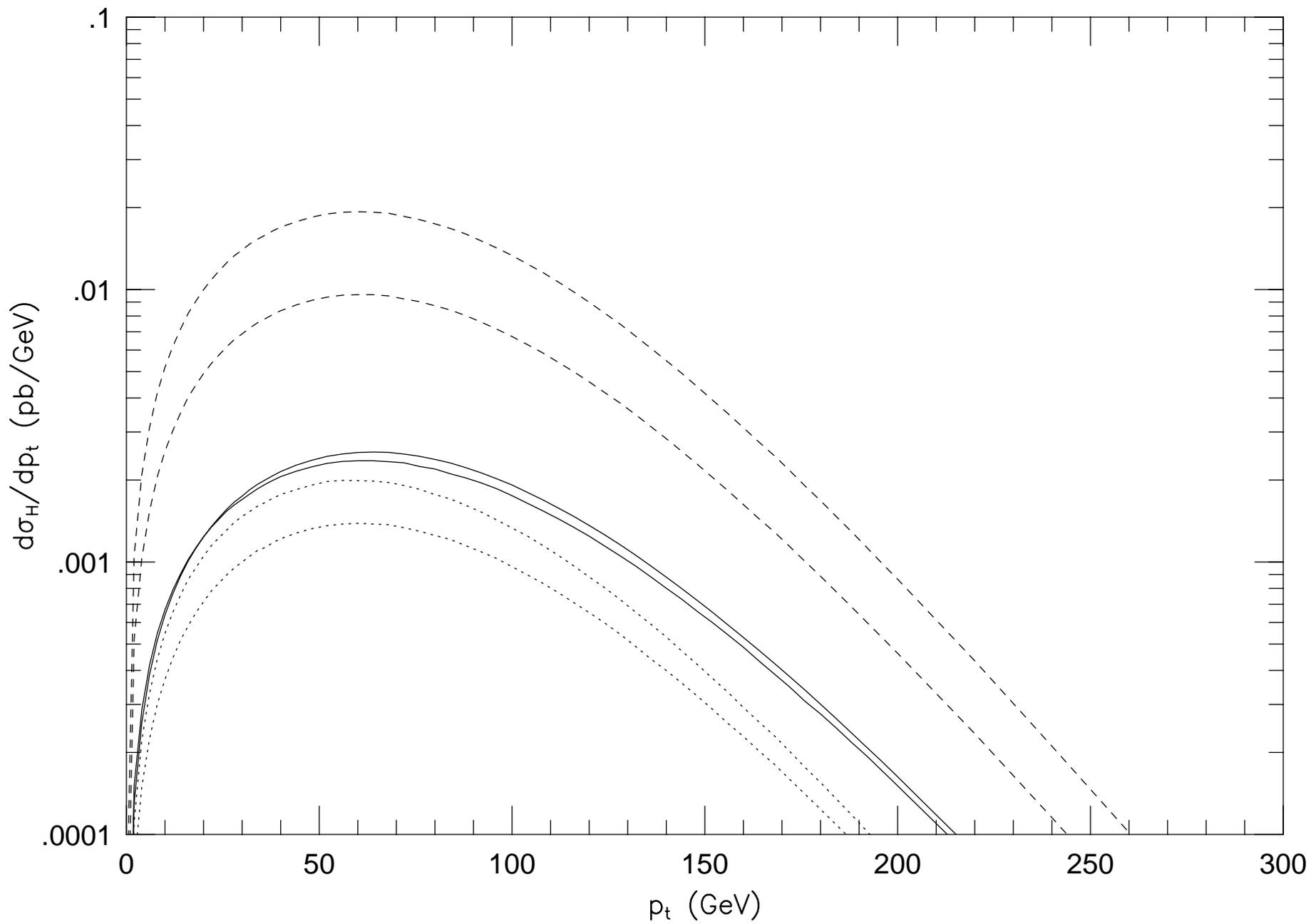
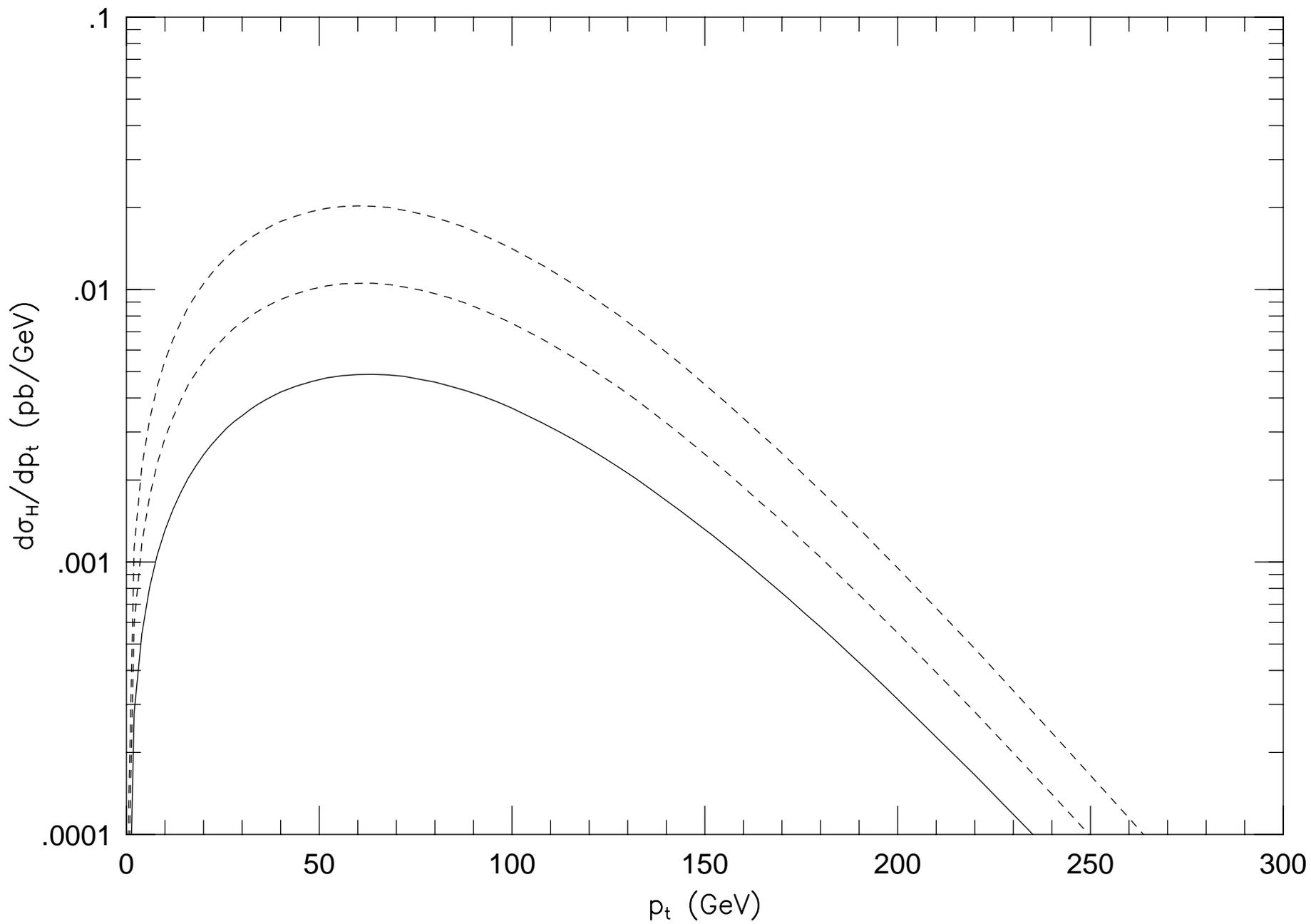
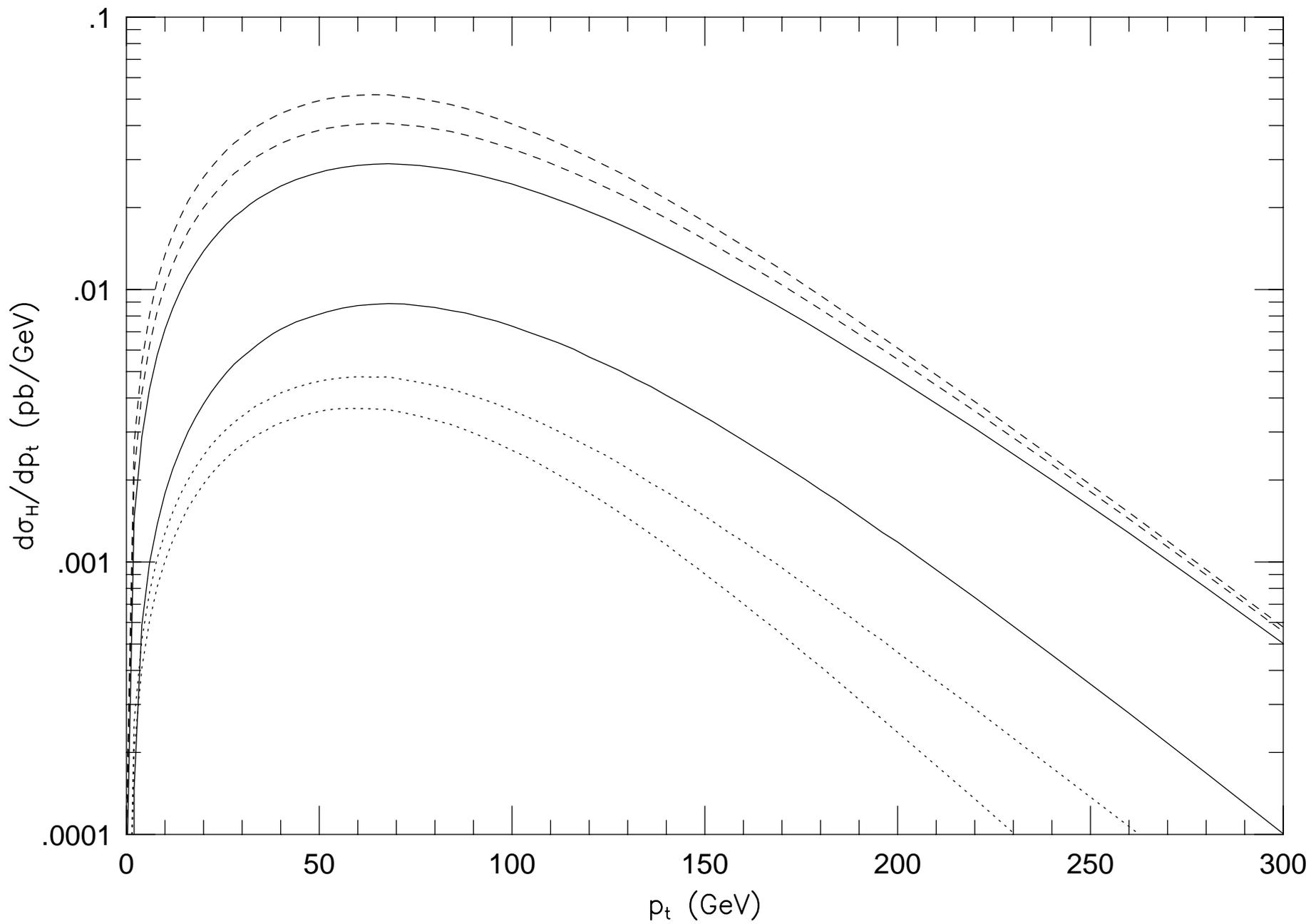
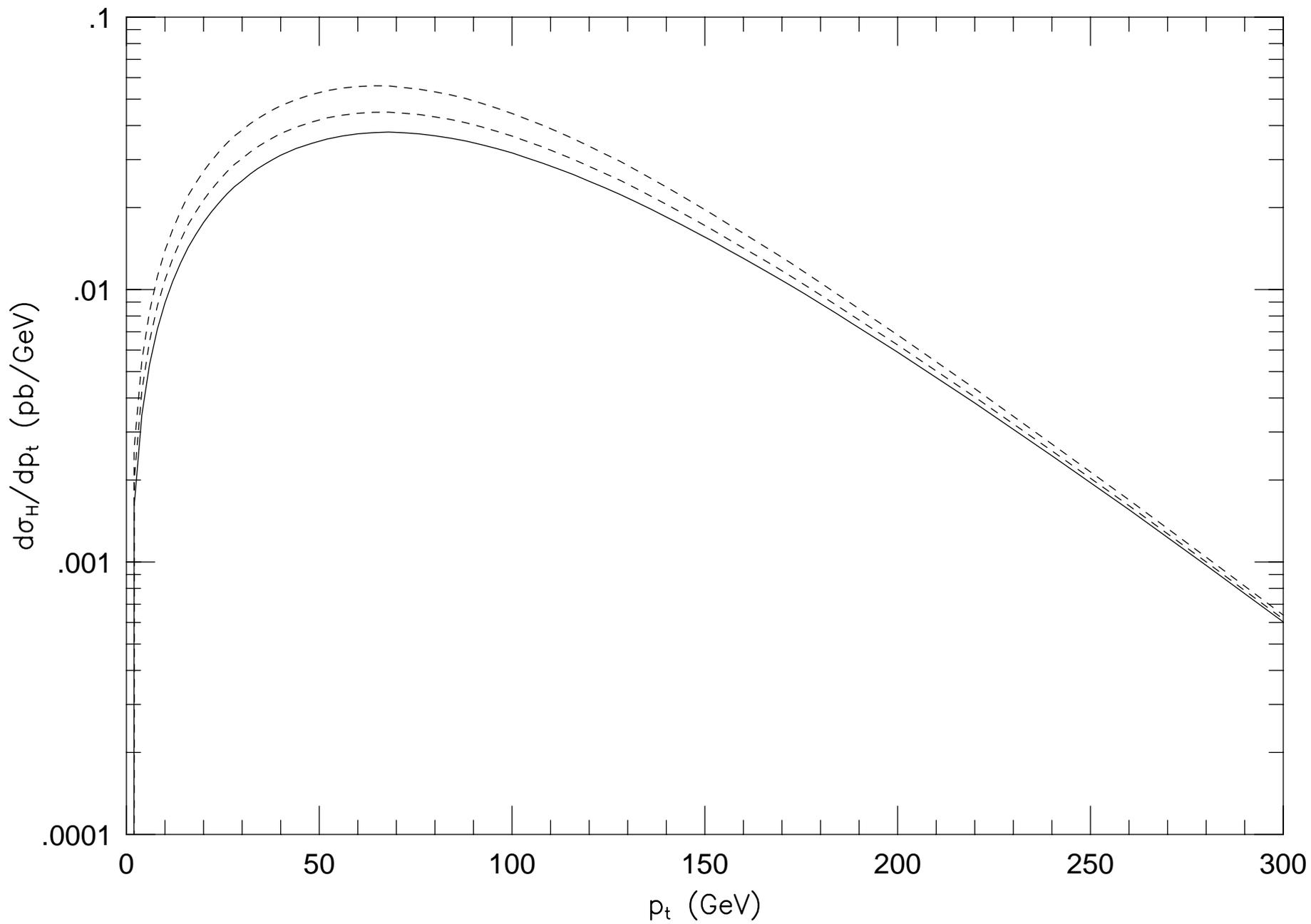
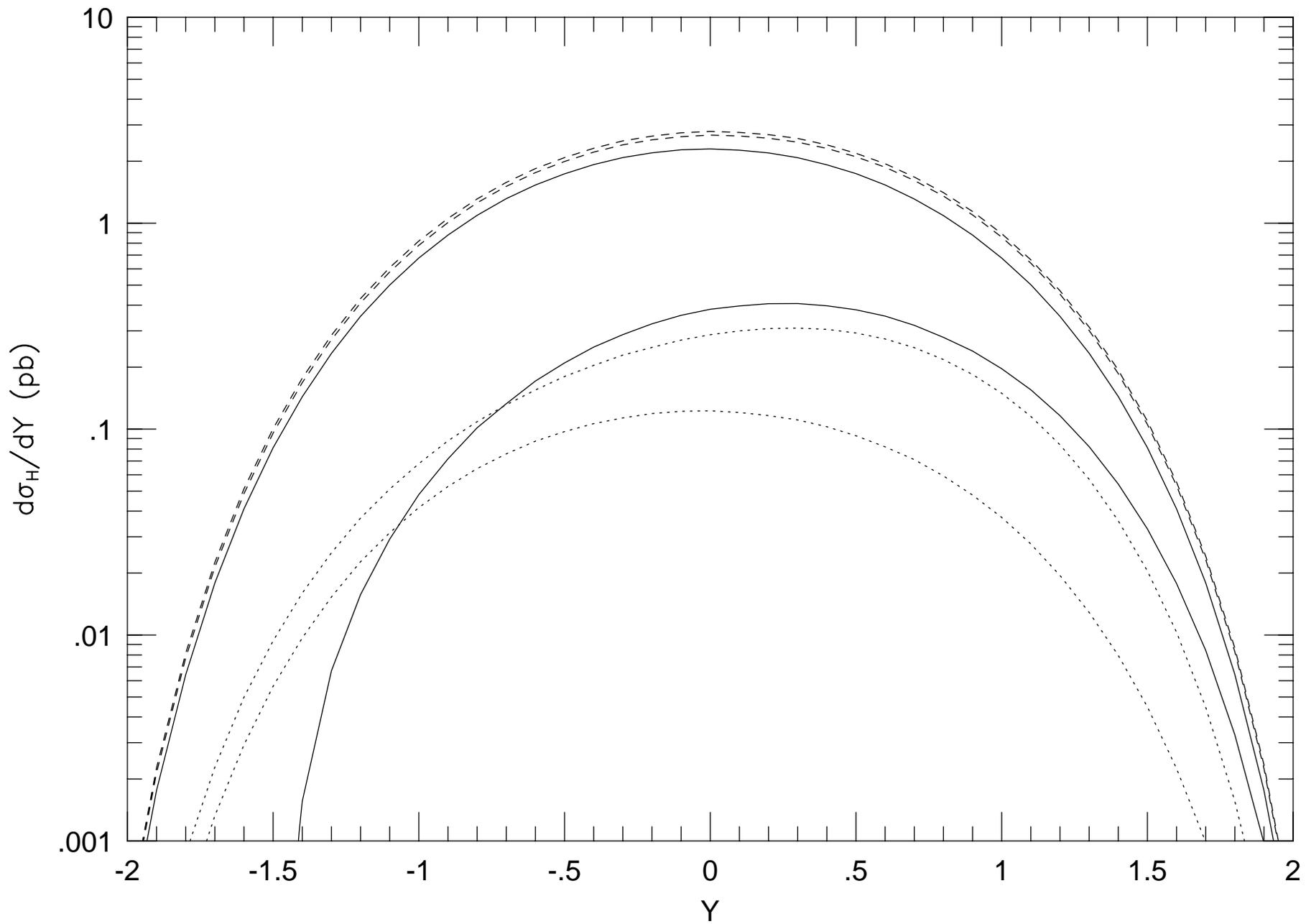
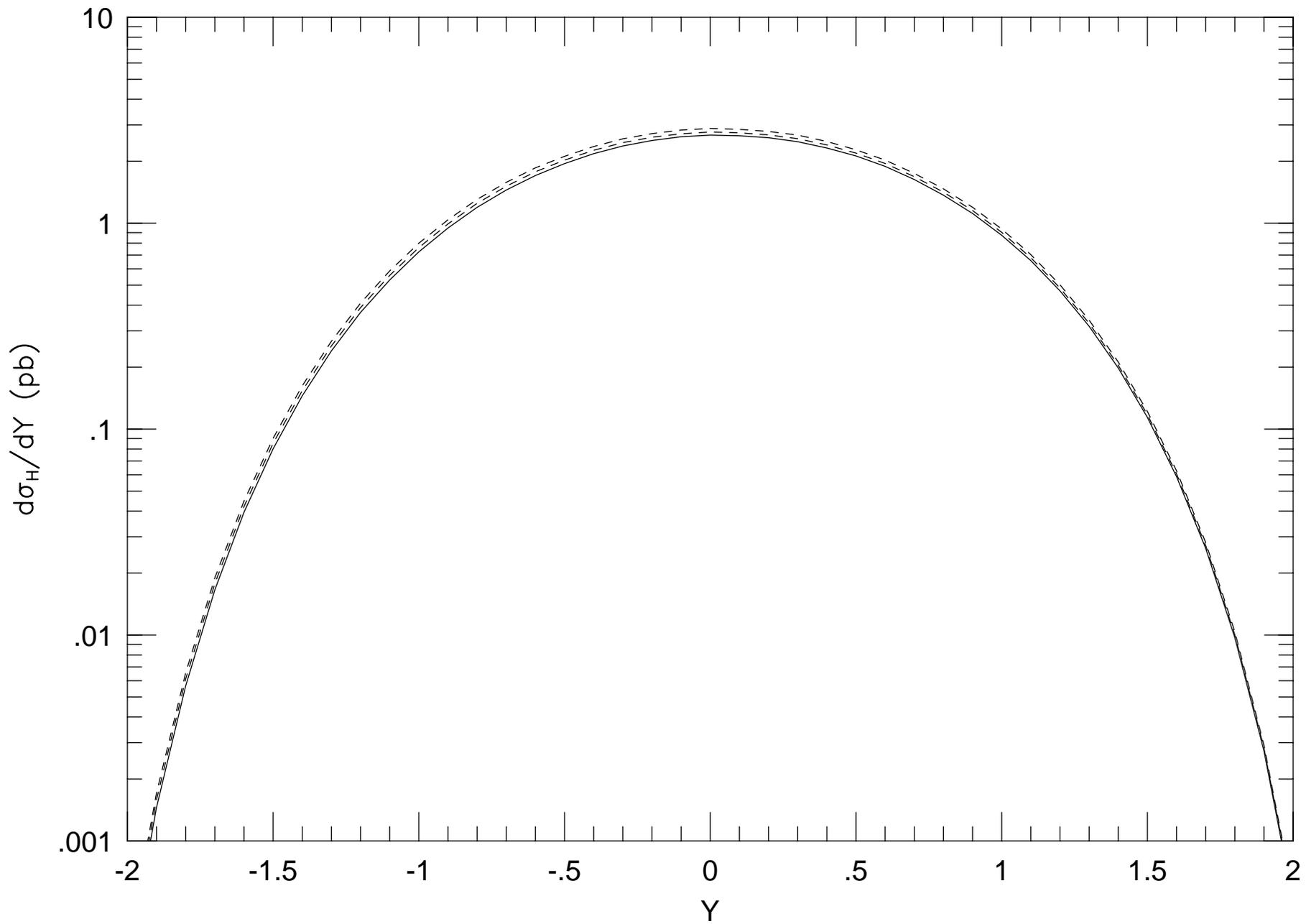
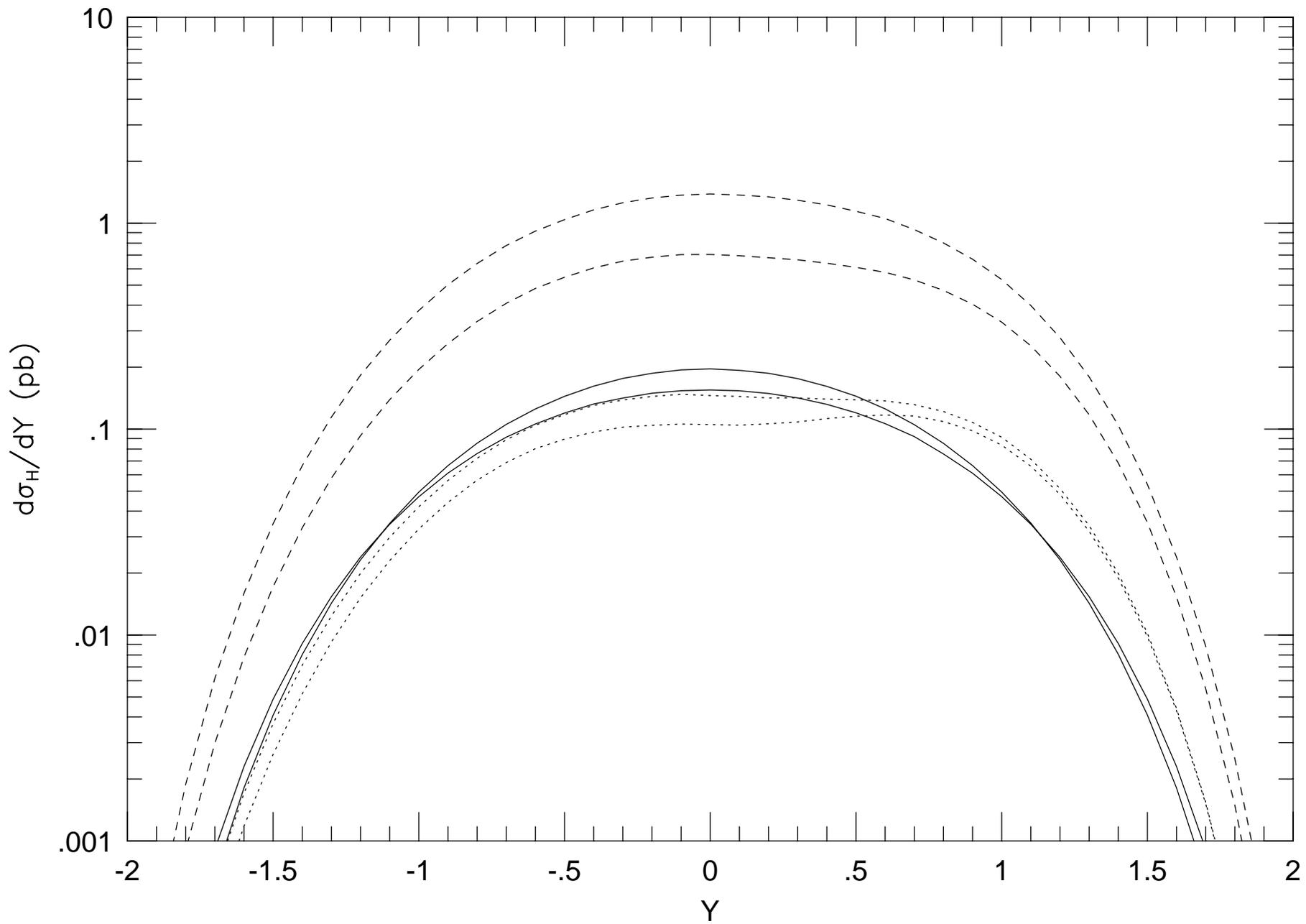
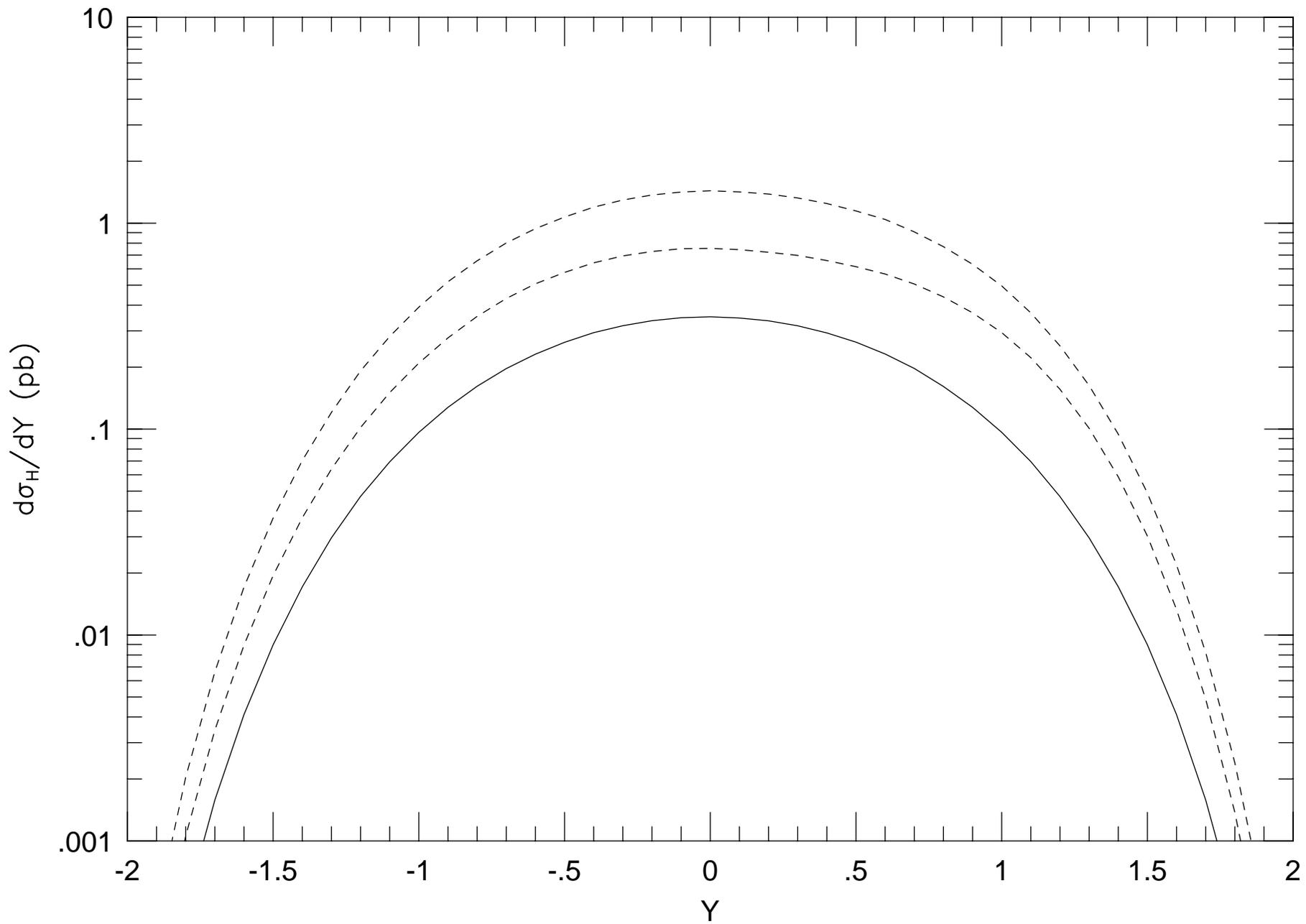
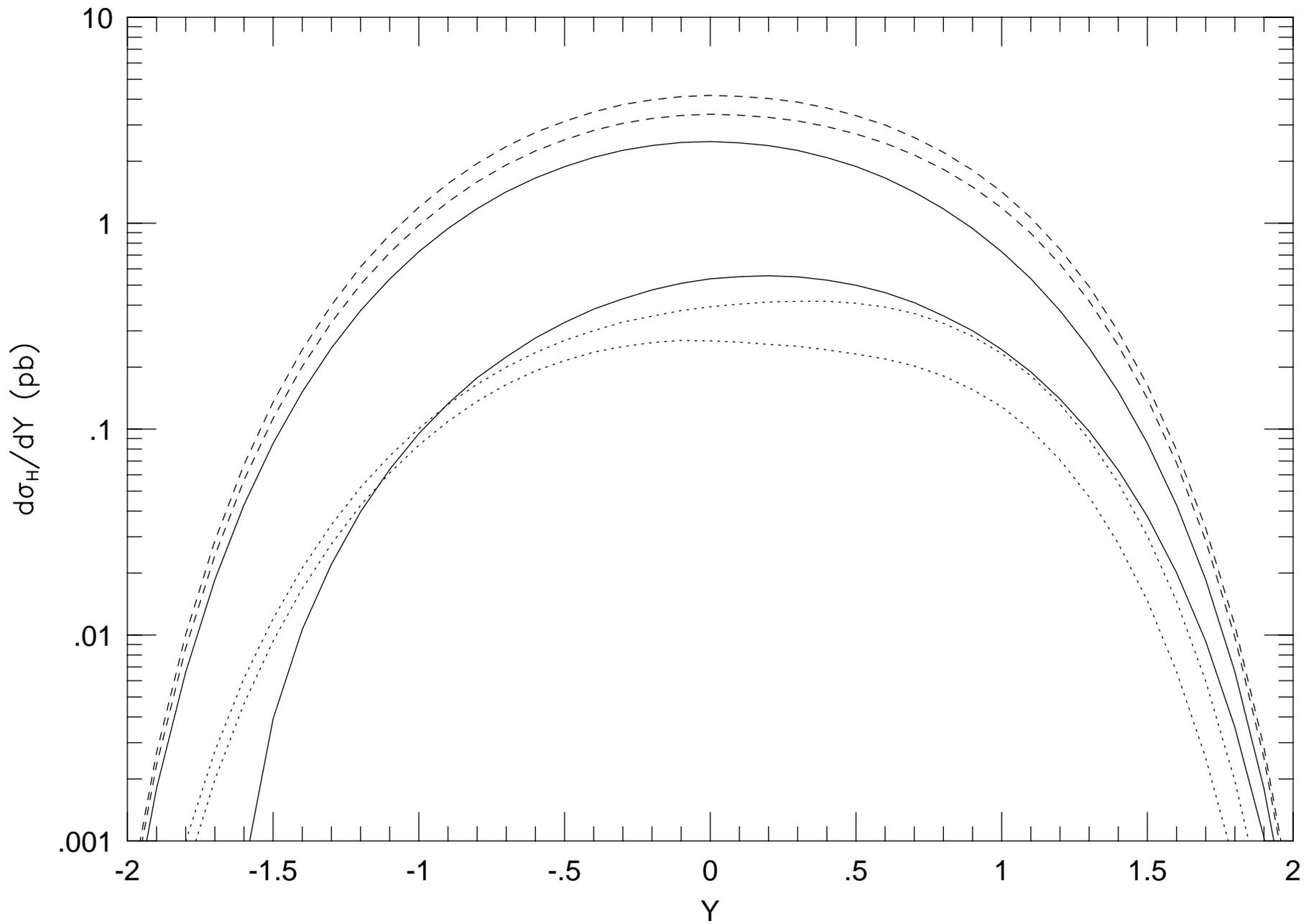
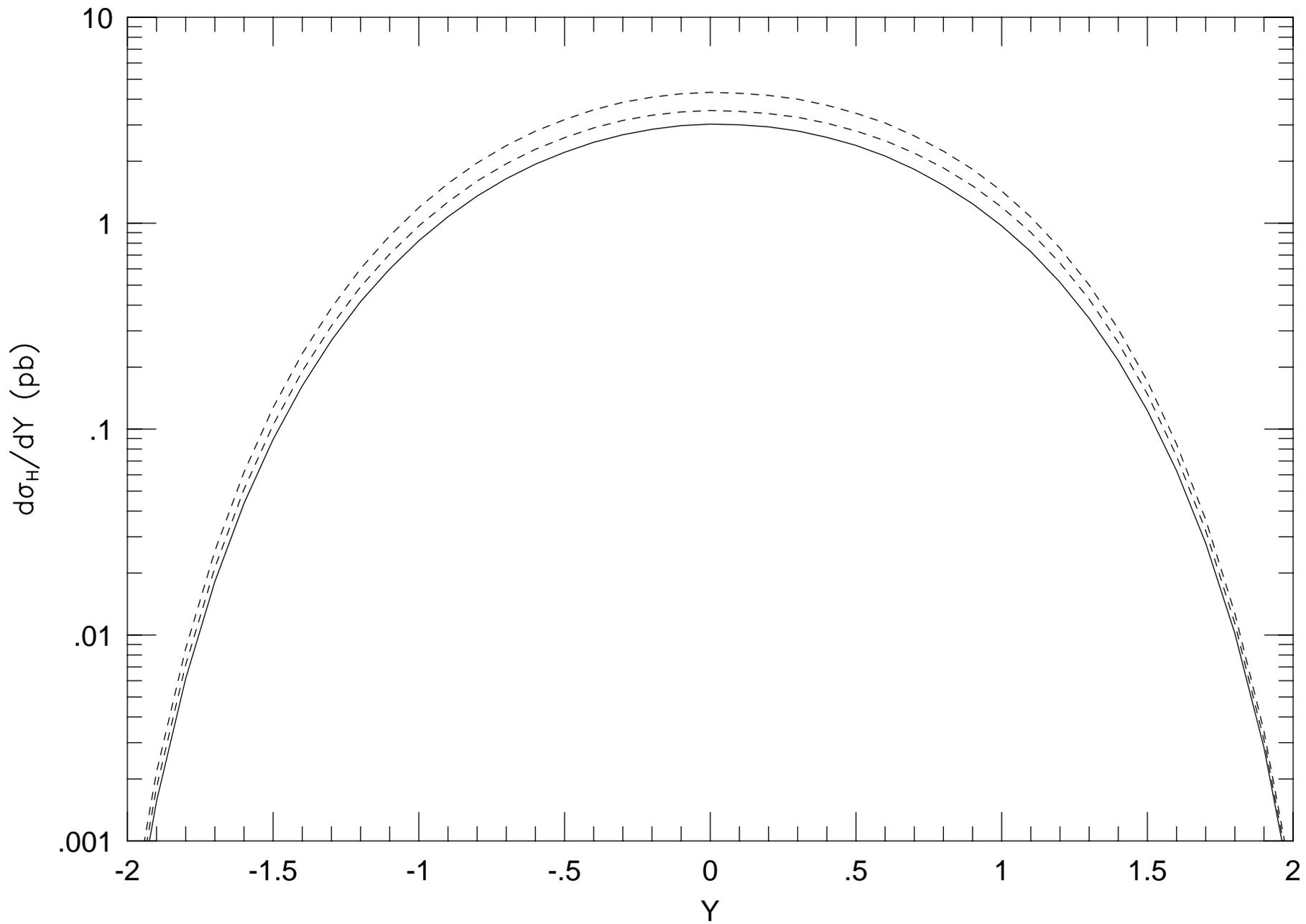
\end{document}